\documentclass[english,prd,twocolumn,superscriptaddress,nofootinbib,preprintnumbers]{revtex4}
\usepackage[latin1]{inputenc}
\usepackage{graphicx}
\usepackage{amssymb}
\usepackage{epstopdf}

\makeatletter

\usepackage{amsfonts}
\usepackage{dcolumn}
\usepackage{hyperref}

\def\be{\begin{equation}}
\def\ee{\end{equation}}
\def\ba{\begin{eqnarray}}
\def\ea{\end{eqnarray}}
\def\bs{\begin{subequations}}
\def\es{\end{subequations}}

\newcommand{\rd}{{\rm d}}

\def \om   {\Omega_{\rm 0m}}

\usepackage{color}

\makeatother

\usepackage{babel}
\makeatother
\begin{document}

\title{The universe is accelerating. Do we need a new mass scale?}

\author{Savvas Nesseris}
\affiliation{The Niels Bohr International Academy, The Niels Bohr
Institute, Blegdamsvej 17, DK-2100, Copenhagen \O, Denmark}
\email{nesseris@nbi.dk}

\author{Federico Piazza}
\affiliation{Perimeter Institute for Theoretical Physics, Waterloo, Ontario, N2L 2Y5, Canada }
\affiliation{Canadian Institute for Theoretical Astrophysics (CITA), Toronto, Canada }
\email{fpiazza@perimeterinstitute.ca}

\author{Shinji Tsujikawa}
\affiliation{Department of Physics, Faculty of Science, Tokyo University of Science,
1-3, Kagurazaka, Shinjuku-ku, Tokyo 162-8601, Japan}
\email{shinji@rs.kagu.tus.ac.jp}

\begin{abstract}

We try to address quantitatively the question whether a new mass
is needed to fit current supernovae data. For this purpose, we
consider an infra-red modification of gravity that does not
contain any new mass scale but systematic subleading corrections
proportional to the curvature. The modifications are of the same
type as the one recently derived by enforcing the ``Ultra Strong
Equivalence Principle" (USEP) upon a
Friedmann-Lema\^{i}tre-Robertson-Walker universe in the presence
of a scalar field. The distance between two comoving observers is
altered by these corrections and the observations at high redshift
affected at any time during the cosmic evolution. While the
specific values of the parameters predicted by USEP are ruled out,
there are regions of parameter space that fit SnIa data very well.
This allows an interesting possibility to explain the apparent
cosmic acceleration today without introducing either a dark energy
component or a new mass scale.

\end{abstract}

\date{\today}

\maketitle

\section{Introduction}

During the last decade, several observational probes
\cite{SNIa,CMB,BAO} have confirmed that our universe is undergoing
a phase of accelerated expansion. Beyond the details of specific
models, one of the most remarkable aspects of such a discovery is
the seemingly unavoidable presence of a new tiny mass scale in the
theory that describes our world.

In the framework of General Relativity (GR), a negative pressure
component (``dark energy''~\cite{review}) can account for the
cosmic acceleration. During the cosmological expansion, such a
component has to become dominant when the average energy density
$\rho(t)$ drops to about its present value $\rho_0$ ($t$ is the
proper time). Thus, dark energy Lagrangians typically contain a
mass parameter of the order of $M \sim \rho_0^{1/4} \sim 10^{-3}
{\rm eV}$, that triggers the epoch when dark energy starts to
dominate\footnote{The mass parameter can be higher e.g. in
quintessence models with power-law
potentials~\cite{Steinhardt:1999nw,Ratra}, but at the price of
giving a milder equation of state which is now severely challenged
by observations.}. Models of massive/modified gravity highlight
the problem from a different perspective. If the graviton is
effectively massive, the modified dynamics of gravity at large
distances can provide a mechanism for ``self-acceleration"
\cite{dgp} and/or of filtering for the cosmological constant's
zero mode~\cite{justin}. In that case the new mass brought into
the theory is the mass of the graviton\footnote{Interestingly, and
as opposed to, e.g., the mass of scalar quintessence fields, such
a mass might be protected against -- and actually made smaller by
-- radiative corrections~\cite{dvali}.}, $m_g$, typically of order
the Hubble constant $H_0$, i.e. $m_g \sim H_0 \sim (\rho_0/M_{\rm
Pl}^2)^{1/2} \sim 10^{-33} {\rm eV}$. In $f(R)$ theories of
gravity, where the Lagrangian density $f$ is a function of the
Ricci scalar $R$, the late-time acceleration is realized, again,
by introducing a curvature scale $R_c$ of the order of
$H_0^2$~\cite{fRviable} (see also Refs.~\cite{fRviable2}). Note
that, although based on different mechanisms, all models follow an
analogous pattern\footnote{The few alternatives to this common
pattern include the proposal that our universe is not homogeneous
on large scales~\cite{inho} and attempts based on possible
non-trivial effects of smaller inhomogeneities on the cosmic
evolution~\cite{back}.}: there is a tuned scale ``hidden" in the
theory which becomes effective, ``by coincidence", when the
appropriate cosmological quantity (the average density $\rho(t)$
or the Hubble parameter $H(t)$) drops to about its value.

But is the acceleration of the universe intrinsically implying the
presence of a new mass scale? If we allow the possibility of
departures from GR at large distances there is a logical
alternative. High redshift observations have the unique property
of relating objects (e.g., the observer and the supernova) that
are placed from each other at a relative distance of the order of
the average inverse curvature (roughly, the Hubble length $\sim
H_0^{-1}$). Therefore, modifying GR in the infrared (IR) at a
length scale set by the curvature -- rather than fixed \emph{a
priori} by a parameter -- will systematically affect any
cosmological observation at high redshift, regardless of when such
an observation takes place and without the need of any external
mass scale. In other words, we might not need a new mass scale
because we already have (a dynamical) one, the Hubble parameter
$H(t)$; the only ``coincidence" that we might be experiencing is
that of observing objects that are placed from us as far as the
Hubble radius\footnote{The same circumstance does not apply, for
instance, to observations within the solar system: typical solar
system distances are always extremely small with respect to, say,
the average Weyl curvature. An order of magnitude estimate indeed
gives $({\rm Weyl\ curvature})^{-1/2} \approx r\,(r/3\,{\rm
km})^{1/2}$, where $r$ is the distance from the Sun.}.

The point of view sketched above is somewhat compelling, it
addresses directly the fine tuning and coincidence problems, but
seems to require a serious revision of the current low energy
framework for gravity. Any gravitational operator that becomes
effective in the infrared, on dimensional grounds, has to bring in
the Lagrangian a mass parameter. Moreover, GR itself is already a
geometrical deformation of flat space at distances of the order of
the curvature. What seems to be required is a further
curvature-dependent subleading effect that systematically modifies
the geometrical description of GR at large distances.

Recently, a proposal along those lines has been made by one of the
present authors. The modification upon the standard framework is
forced by imposing an ``Ultra Strong" version of the Equivalence
Principle (USEP, see Refs.~\cite{Piazza1,fedo} for more details).
USEP suggests that the usual geometric description of spacetime as
a metric Riemannian manifold might hold only approximately, at
small distances. Such a conjectured ``IR-completion" of gravity,
in its full generality, represents a major theoretical challenge.
However, it can be tentatively explored with a Taylor expansion
around GR, by applying USEP to a specific GR solution\footnote{In
a similar fashion, someone who does not know GR can try to expand
around some point in Riemann normal coordinates and find, in some
specific cases, the first GR corrections to flat space.} (see
Appendix \ref{sketch}). For a spectator scalar field in a
spatially flat Friedmann-Lema\^{i}tre-Robertson-Walker (FLRW)
universe, the first-order correction to GR is calculable, and few
cosmological consequences are derivable~\cite{fedo}.

Consider, as the  zeroth-order (GR-) approximation, a homogeneous,
spatially flat FLRW universe with scale factor $a(t)$. It is known
that in such a solution the physical distance $d(t)$ between a
pair of comoving observers grows proportionally to $a(t)$;
otherwise stated, the comoving distance $\lambda \equiv d(t)/a(t)$
is a constant. The first-order correction found in
Ref.~\cite{fedo} modifies such an expansion law by a subleading
distance-dependent contribution. As a result, the distance $d(t)$
between two comoving observers grows as $a(t)$ only when  small
compared to the Hubble length $H^{-1}$ but gets relevant
corrections otherwise. Thus, the scale factor $a(t)$ defines the
expansion everywhere but only in the local limit, and effectively
detaches from the expansion on the largest scales.

The corrected expansion is most easily seen in terms of the above
defined comoving distance $\lambda$, which is constant only in the
small distance limit. Its derivative with respect to observers'
proper time $t$ reads~\cite{fedo}
\begin{equation}
\dot{\lambda}=\lambda^3 \frac{(H^2 a^2)^{\cdot}}{4}\,+ {\rm higher\ orders},
\label{dotlamre}
\end{equation}
and is clearly negligible on sub-Hubble scales. For completeness,
a basic derivation of Eq.~(\ref{dotlamre}) is sketched in
Appendix~\ref{sketch}. The comoving trajectory $r(t)$ of a light
ray also receives corrections because the modified global
expansion (\ref{dotlamre}) has to be considered on top of the
usual contribution $\rd r = \rd t /a$. In a matter-dominated
universe, where the Hubble parameter at the redshift $z=1/a-1$ is
given by $H(z)=H_0(1+z)^{3/2}$, we have
\begin{equation}
\frac{\rd (H_0 r)}{\rd z}=\frac{1}{(1+z)^{3/2}}
+\frac{(H_0 r)^3}{4}\,,
\label{redshift}
\end{equation}
which has to be solved with initial conditions $r(0) = 0$.

The above modification, including the factor of $1/4$, is forced
by requiring that USEP applies for a scalar field in a spatially
flat FLRW universe~\cite{fedo}. The correction increases the
luminosity distance
\begin{equation}
d_L(z)=(1+z)r(z)\,,
\label{luminosity}
\end{equation}
and therefore it effectively goes in the direction of a universe
with positive acceleration. However, as the present analysis shows
as a by-product, the correction given in the last term of
Eq.~(\ref{redshift}) is too small (of too high order in the
redshift) to explain SnIa data.

Equation (\ref{redshift}) is an example of a IR-geometrical
deformation that does not contain any mass scale but only
$H$-dependent subleading terms. In this paper, by studying a
generalized version of (\ref{redshift}), we attempt to address
quantitatively the question whether SnIa data can be explained
without introducing any new mass scale. We include terms of lower
order in the redshift that are needed to efficiently reproduce
SnIa observations.  Equation (\ref{dotlamre}) is generalized as
follows:
\begin{eqnarray}
\dot \lambda & = & A_1\,  \lambda H + A_2 \,  \lambda^2 H^2 a + \dots \nonumber \\
& & +B_1 \, \lambda^2 (H a)^{\cdot} + B_2 \, \lambda^3  (H^2 a^2)^{\cdot} + \dots\, .
\label{dotlam}
\end{eqnarray}
Note that the above structure of corrections includes
(\ref{dotlamre}) as a special case. In practice, the terms in the
above expansion rearrange when we calculate the luminosity
distance. Therefore, for a matter-dominated universe, a quite
general structure of subleading terms is given by
\begin{equation} \label{ODE}
\frac{\rd (H_0 r)}{\rd z} = \frac{1}{(1+z)^{3/2}} F\left(H_0 r (1+z)^{1/2}\right),
\end{equation}
where $F(x)$ is a generic function with $F(0) = 1$:
\begin{equation} \label{ODE2}
F(x) = 1 + \alpha x + \beta x^2 + \gamma x^3+\, \dots \, .
\end{equation}
By comparison with (\ref{dotlam}) we have $ \alpha=-A_1,
\beta=B_1/2-A_2, \gamma=B_2\,.$ Note that Eq.~(\ref{redshift})
corresponds to $\alpha = \beta = 0$ and $\gamma=1/4$, while in GR
all coefficients are set to zero.

\section{Effective description}

It is possible to obtain analytic solutions of (\ref{ODE}) in some
restricted cases (e.g., $\gamma=0$, see Appendix
\ref{appendixsec}). However, it is perhaps more useful to study
the effective behavior of (\ref{ODE}) at low $z$. In order to make
an easy comparison with known parameterizations of dark energy, we
can easily express our first two parameters, $\alpha$ and $\beta$,
in terms of an effective density parameter $\Omega_{\rm DE}^{\rm
eff}$ and a {\it constant} effective equation of state $w_{\rm
eff}$ of dark energy, in the presence of non-relativistic
matter\footnote{We can do a similar exercise for an evolving
effective equation of state $w_{\rm eff}(z)$ instead of constant
$w_{\rm eff}$. However the corresponding expression in this case
is much more complicated, so we will not present it here.}. Such
effective parameters~\cite{luca} are thus defined by
\begin{equation}
r(z)=\int_0^z\frac{1}{H_{\rm eff}(x)}\,{\rm d}x\,,
\label{rzreal}
\end{equation}
where
\begin{equation}
H_{\rm eff}(z)\equiv H_0 \sqrt{(1 - \Omega_{\rm DE}^{\rm
eff})(1+z)^3+ \Omega_{\rm DE}^{\rm eff} (1+z)^{3 (1+ w_{\rm eff})}}.
\label{heff}
\end{equation}
By expanding Eq.~(\ref{rzreal}) at small redshift we find
\begin{eqnarray}
H_0 r &=& z - \label{series2}
\frac{3 z^2}{4}\left(1 + \Omega_{\rm DE}^{\rm eff} w_{\rm eff} \right)
\nonumber  \\
&& +\frac{z^3}{8} \left[5 + 8w_{\rm eff}\Omega_{\rm DE}^{\rm eff}
+3w_{\rm eff}^2 \Omega_{\rm DE} (3\Omega_{\rm DE}-2)
\right] \nonumber \\
& &+\, \dots \,.
\end{eqnarray}
On the other hand, the solution of (\ref{ODE}) can be expanded as
\begin{eqnarray} \label{series1}
H_0 r &=&  z - \frac{3 z^2}{4}\left(1 -
\frac{2}{3}\alpha\right) \nonumber \\
&&+ \frac{z^3}{24} \left(15 -
14\alpha + 4 \alpha^2 + 8 \beta\right)+\, \dots \,.
\end{eqnarray}

By comparing~(\ref{series2}) and~(\ref{series1})
we can relate the two sets of parameters:
\begin{eqnarray} \label{params}
\alpha &=& -\frac{3}{2} w_{{\rm eff}} \Omega_{\rm DE}^{\rm eff},
\nonumber \\
\beta &=& \frac{3}{8} w_{{\rm eff}} \left[ 1+6 w_{{\rm eff}}
\left( \Omega_{\rm DE}^{\rm eff}-1 \right)\right]
\Omega_{\rm DE}^{\rm eff}\,.
\end{eqnarray}
For the $\Lambda$CDM model ($w_{\rm eff}=-1$) with $\Omega_{\rm
DE}^{\rm eff}=0.7$ it follows that $\alpha=1.05$, $\beta=-0.74$.
Of course the above expansions are valid only for $z \ll 1$, so it
is expected that the likelihood analysis including high-redshift
data can give different constraints on the model parameters (as we
will see later).

The most important correction that leads to a larger comoving
distance relative to the Einstein de Sitter universe originates
from the $\alpha$ term in Eqs.~(\ref{ODE})-(\ref{ODE2}), i.e., the
term $A_1 \lambda H$ in Eq.~(\ref{dotlam}). Note that, for $z \ll
1$, the correction $\gamma$ becomes important only for the terms
higher than order $z^3$ in Eq.~(\ref{series1}). Hence, we expect
that Eq.~(\ref{redshift}) alone will not be sufficient to
reproduce SnIa data efficiently, at least at low redshift.

\section{The $\textrm{SnIa}$ data analysis}
\label{SnIa-analysis}

In this section we shall present a method to place observational
constraints on the IR corrections (\ref{ODE})-(\ref{ODE2}) from
SnIa data. We will use the SnIa dataset of Hicken {\it et
al.}~\cite{Hicken:2009dk} consisting in total of 397 SnIa out of
which 100 come from the new CfA3 sample and the rest from Kowalski
{\it et al.}~\cite{Kowalski:2008ez}. These observations provide
the apparent magnitude $m_{\rm th}(z)$ of the SnIa at peak
brightness after implementing the correction for galactic
extinction, the K-correction and the light curve width-luminosity
correction. The resulting apparent magnitude $m_{\rm th}(z)$ is
related to the luminosity distance $d_L(z)=(1+z)r(z)$ through
\begin{equation}
m_{\rm th}(z)={\bar M} (M,H_0) + 5\,{\rm log}_{10} (d_L (z))\,,
\label{mdl}
\end{equation}
where ${\bar M}$ is the magnitude zero point offset and depends on
the absolute magnitude $M$ and on the present Hubble
parameter $H_0$ as
\begin{equation}
{\bar M} =M + 5\,{\rm log}_{10}\left(\frac{H_0^{-1}}{{\rm Mpc}}\right)
+ 25=M-5\,{\rm log}_{10}h+42.38.
\label{barm}
\end{equation}
Here the absolute magnitude $M$ is assumed to be constant
after the above mentioned corrections have been
implemented in $m_{\rm th}(z)$.

The SnIa datapoints are given, after the corrections have been
implemented, in terms of the distance modulus
\begin{equation}
\mu_{\rm obs}(z_i)\equiv m_{\rm obs}(z_i) - M\,.
\label{mug}
\end{equation}
The theoretical model parameters are determined by
minimizing the quantity
\begin{equation}
\chi^2_{\rm SnIa}=\sum_{i=1}^N
\frac{[\mu_{\rm obs}(z_i) - \mu_{\rm th}(z_i)]^2}
{\sigma_{\mu \; i}^2}\,,
\label{chi2}
\end{equation}
where $N=397$, and $\sigma_{\mu \; i}^2$ are
the errors due to flux uncertainties, intrinsic dispersion of SnIa
absolute magnitude and peculiar velocity dispersion.
These errors are assumed to be Gaussian and uncorrelated.
The theoretical distance modulus is defined as
\begin{equation}
\mu_{\rm th}(z_i)\equiv m_{\rm th}(z_i)
- M =5\,{\rm log}_{10} (d_L (z)) +\mu_0\,,
\label{mth}
\end{equation}
where $\mu_0= 42.38 - 5 \,{\rm log}_{10}h$ and $\mu_{\rm obs}$ is
given by Eq.~(\ref{mug}). The steps we have followed for the minimization
of Eq.~(\ref{chi2}) in terms of its parameters are described in detail
in Refs.~\cite{Nesseris:2004wj,Nesseris:2005ur,Nesseris:2006er}.

We will also use the two information criteria known as AIC (Akaike
Information Criterion) and BIC (Bayesian Information Criterion),
see Ref.~\cite{Liddle:2004nh} and references there in. The AIC is
defined as
\be
{\rm AIC} =-2\ln \mathcal{L}_{\rm max}+2k\,,
\label{aic}
\ee
where the likelihood is defined as $\mathcal{L}\propto
e^{-\chi^2/2}$, the term $-2\ln \mathcal{L}_{\rm max}$
corresponds to the minimum $\chi^2$ and $k$ is the number
of parameters of the model. The BIC is defined similarly as
\be
{\rm BIC}=-2\ln
\mathcal{L}_{\rm max}+k \ln\,N\,,
\label{bic}
\ee
where $N$ is the number of
datapoints in the set under consideration. According to these
criteria a model with the smaller AIC/BIC is considered to be the
best and specifically, for the BIC a difference of 2 is considered
as positive evidence, while 6 or more as strong evidence in favor
of the model with the smaller value. Similarly, for the AIC a
difference in the range $0 - 2$ means that the two models have
about the same support from the data as the best one, for a
difference in the range $2 - 4$ this support is considerably less
for the model with the larger AIC, while for a difference $> 10$
the model with the larger AIC practically irrelevant
\cite{Liddle:2004nh,Biesiada:2007um}.

\begin{figure*}[!t]
\centering
\vspace{0cm}\rotatebox{0}{\vspace{0cm}\hspace{0cm}\resizebox{.48\textwidth}{!}{\includegraphics{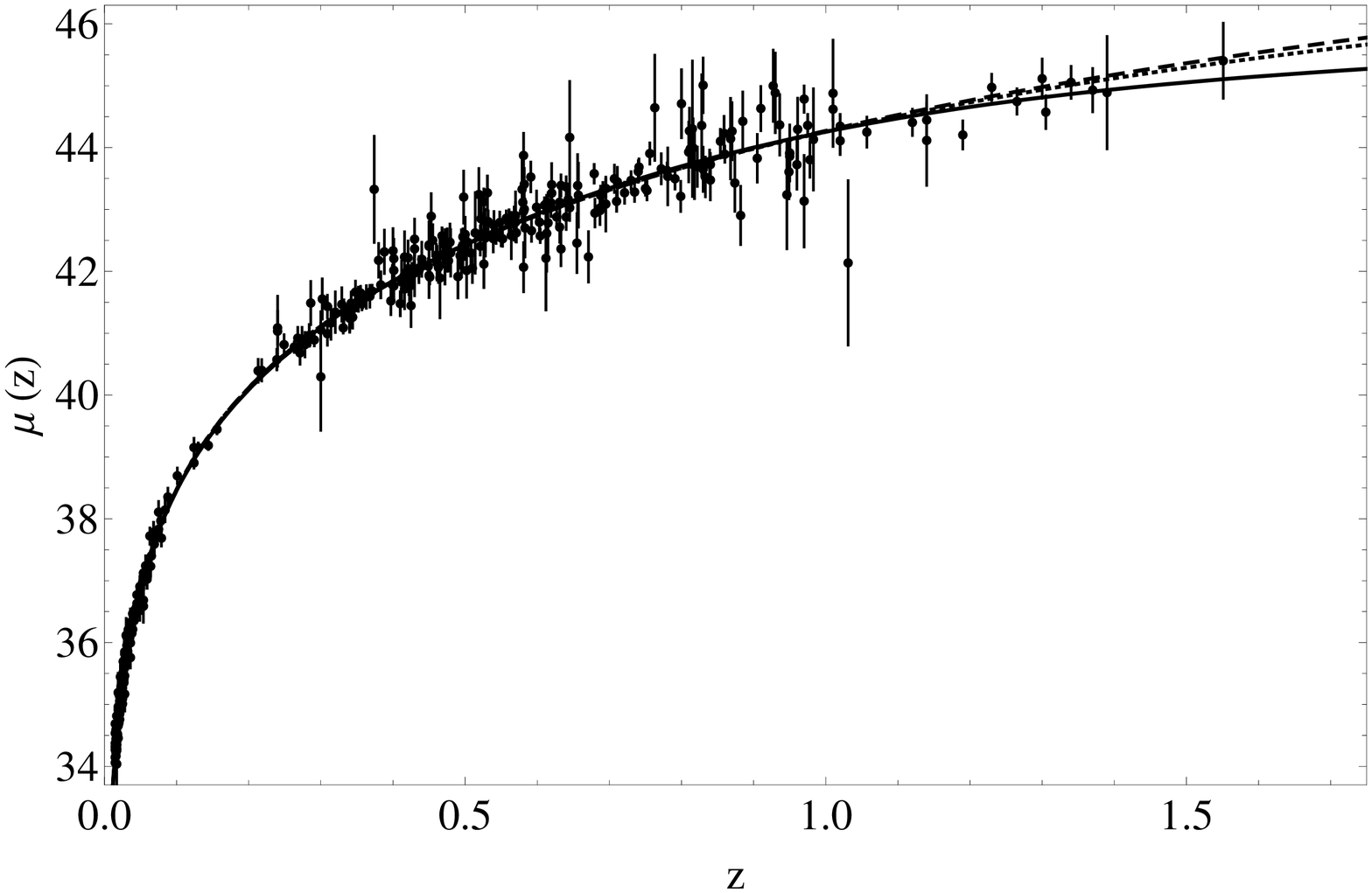}}}
\vspace{0cm}\rotatebox{0}{\vspace{0cm}\hspace{0cm}\resizebox{.49\textwidth}{!}{\includegraphics{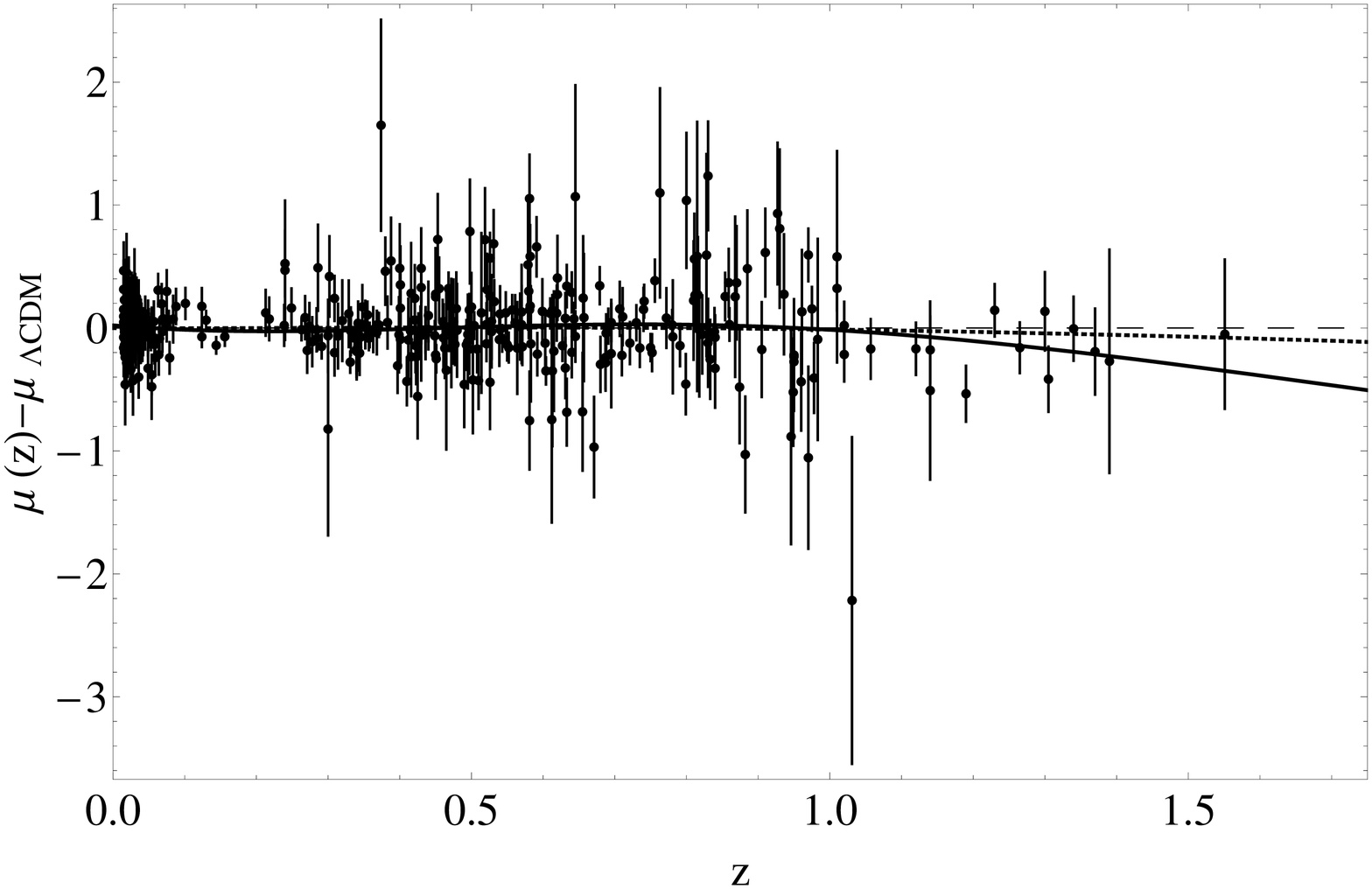}}}
\caption{Left: The distance modulus for the best fit $\Lambda$CDM
model (dashed line) and the SCDM model + IR correction of
Eq.~(\ref{ODE}) with the 3 parameter case (solid black line) and
the 2 parameter case with $\gamma=0$ (dotted line). Right: The
residuals relative to $\Lambda$CDM for the model with the IR
correction of Eq.~(\ref{ODE}) with the 3 parameter case (solid
black line) and the 2 parameter case with $\gamma=0$ (dotted
line). The dashed line corresponds to zero.\label{plot1}}
\end{figure*}

\begin{figure}[!t]
\centering
\vspace{0cm}\rotatebox{0}{\vspace{0cm}\hspace{0cm}\resizebox{.48\textwidth}{!}{\includegraphics{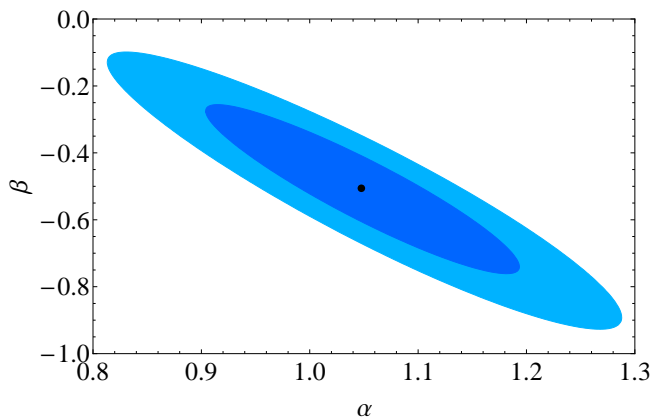}}}
\caption{The 1$\sigma$ and 2$\sigma$ contours in the $(\alpha,
\beta)$ plane for the 2 parameter model with $\gamma=0$. The black
dot ($\alpha=1.05$, $\beta=-0.51$) corresponds to the best fit.
\label{plot2}}
\end{figure}

\section{Results}
\label{results}

We solve Eq.~(\ref{ODE})-(\ref{ODE2}) numerically to find $r(z)$
for the matter-dominated model (SCDM).
Then the model is tested against the SnIa data by using
Eqs.~(\ref{luminosity}), (\ref{chi2}), and (\ref{mth}).
Since the parameter $\gamma$ does not appear up to third order
in the expansion of Eq.~(\ref{series1}), we will also consider
the case where $\gamma=0$.

In Fig.~\ref{plot1} (left) we present the best fit distance
modulus versus the redshift $z$ for the best fit $\Lambda$CDM
model (dashed line), with the present matter density parameter
$\om=0.289^{+0.023}_{-0.022}$) and the SCDM model + IR correction
of Eq.~(\ref{series1}) with all 3 parameters (solid black line)
and the 2 parameter case with $\gamma=0$ (dotted line). For the
two parameter case we find that $\alpha= 1.05^{+0.15}_{-0.14}$ and
$\beta=-0.51^{+0.23}_{-0.23}$ for a $\chi^2= 464.031$ or a
$\chi^2$ per degree of freedom $\sim 1.17$, whereas the best fit
$\Lambda$CDM has a $\chi^2$ per degree of freedom $\sim 1.18$.

In Fig.~\ref{plot1} (right) we plot the residuals of our best fit
relative to the $\Lambda$CDM model, i.e. the difference of
$\mu_{\rm best fit}(z)-\mu_{\Lambda{\rm CDM}}(z)$, for the model
with the IR correction of Eq.~(\ref{ODE}) with all 3 parameters
(solid black line) and the 2 parameter case and $\gamma=0$ (dotted
line). In Fig.~\ref{plot2} we show the 1$\sigma$ and 2$\sigma$
contours for the parameters $\alpha$ and $\beta$ of
Eq.~(\ref{ODE}) with $\gamma=0$. The black dot indicates the
best-fit.

We should note that the two parameter model fits very well the
data even if the exact numbers $\alpha=1$, $\beta=-1/2$ and
$\gamma=0$ are used. In this case we find that $\chi^2=465.462$
and a $\chi^2$ per degree of freedom $\sim 1.18$, which is the
same as the best fit $\Lambda$CDM model. The reason for this
success of the model can be seen from the first of
Eq.~(\ref{params}). The standard cosmological model predicts
values of $w_{\rm eff} \simeq -1$ and $\Omega_{\rm DE}^{\rm eff}
\simeq 0.7$, which gives $\alpha \simeq 1$. The value of $\beta
\simeq -0.7$ derived from the second of Eqs.~(\ref{params}) is
slightly different from its best fit $\beta \simeq -0.5$ because
of the limitation of the expansion (\ref{series1}) valid only in
the low redshift regime.

The original model of Ref.~\cite{fedo}, i.e. Eq.~(\ref{redshift}),
corresponds to $\alpha = \beta = 0$, $\gamma = 1/4$. In this case,
however, the agreement is not very good as the $\chi^2$ per degree
of freedom is $\sim 2.21$ with the difference from the
$\Lambda$CDM model being about $20 \sigma$. This discrepancy can
be explained by the fact that the IR correction does not kick in
early enough to allow for good compatibility with the data, while
in low redshifts the SCDM behavior of the model dominates. This
latter property comes from the fact that the $\gamma$-dependent
term appears only at the order of $z^4$ in Eq.~(\ref{series1}).

If we consider all three parameters $\alpha$, $\beta$ and $\gamma$
to be free, then the best fit parameters are $\alpha=0.61$,
$\beta=1.29$ and $\gamma=-1.51$ for a $\chi^2=459.424$ or a
$\chi^2$ per degree of freedom $\sim 1.17$. As it can be seen in
Fig.~\ref{plot1} (solid black line), the corresponding luminosity
distance shows a more significant departure from $\Lambda$CDM at
high redshift.

Finally, it is interesting to consider the case where we fix the
parameters $\alpha$ and $\beta$ to the exact numbers $\alpha=1$,
$\beta=-1/2$ and allow $\gamma$ to vary. In this case we expect
that by changing $\gamma$ we will be able to improve $\chi^2$ and
still be able to compare with $\Lambda$CDM as there will be only
one free parameter in the model. The result is $\chi^2=465.082$
for $\gamma = 0.052$, which is slightly worse ($\delta\chi^2 \sim
1$) than the three parameter case, but it is slightly better
($\delta\chi^2 \sim 0.5$) than $\Lambda$CDM ($\chi^2=465.513$ for
$\om = 0.289$).

\vspace{0pt}
\begin{table}[!b]
\begin{center}
\caption{Comparison of the one, two and three parameter models to
$\Lambda$CDM. Note that the differences  for the AIC and BIC are
in both cases with regard to the model with the minimum value for
the corresponding criterion. For the definition of the AIC and
BIC, see Eqs.~(\ref{aic}) and (\ref{bic}).\label{criteria}}
\begin{tabular}{cccccc}
\hline
\hline\\
\hspace{6pt}\vspace{6pt}   Model     & \hspace{6pt}AIC   & \hspace{6pt}$\Delta$AIC    & \hspace{6pt}BIC   &\hspace{6pt}$\Delta$BIC \\
\hline \\
\hspace{6pt}\vspace{6pt}  $\Lambda$CDM  &\hspace{6pt} $467.513$ &\hspace{6pt} $2.089$  &\hspace{6pt} $471.497$  &\hspace{6pt} $0.000$  \\
\hspace{6pt}\vspace{6pt}  1 parameter   &\hspace{6pt} $467.082$ &\hspace{6pt} $1.658$  &\hspace{6pt} $471.066$  &\hspace{6pt}$-0.431$  \\
\hspace{6pt}\vspace{6pt}  2 parameters  &\hspace{6pt} $468.031$ &\hspace{6pt} $2.607$  &\hspace{6pt} $475.999$  &\hspace{6pt} $4.502$  \\
\hspace{6pt}\vspace{6pt}  3 parameters  &\hspace{6pt} $465.424$ &\hspace{6pt} $0.000$  &\hspace{6pt} $477.376$  &\hspace{6pt} $5.879$  \\
\hline \hline
\end{tabular}
\end{center}
\end{table}

In Table \ref{criteria} we present the results of the two
information criteria, the AIC and the BIC. According to the AIC
the three parameter model is the best, with the $\Lambda$CDM
having considerably less support, as the difference between the
two is slightly larger than 2. On the contrary, using the BIC we
find that the $\Lambda$CDM is now the best and also having
positive evidence against the other models. In all cases, the two
parameter model fairs moderately with both the AIC and the BIC.

\section{Conclusions}
\label{conclusions}

We have considered IR modifications of gravity that do not imply
the presence of a new mass scale in the theory and we have studied
their compatibility with the SnIa data. Our first result is that
the mechanism derived in Ref.~\cite{fedo} (see also Appendix
\ref{sketch}), Eq.~(\ref{dotlamre}), is not enough, by itself, to
describe the observed amount of acceleration. The absence of free
parameters in equation (\ref{dotlamre}) (the model in~\cite{fedo}
has one parameter less than $\Lambda$CDM) does not make up
for the very poor fit of the data. However, a more general
structure of corrections (\ref{dotlam}) can lead to a sensibly
larger luminosity distance than in the Einstein de Sitter
universe. In particular, when $\gamma=0$ in
Eq.~(\ref{ODE})-(\ref{ODE2}), we have found that the model fits
the data very well for the values close to the exact numbers
$\alpha=1$ and $\beta=-1/2$. It is interesting to consider such sharp
numerical values, not because of abstract numerology, but because
a mechanism analogous to that described in Appendix~\ref{sketch}
very naturally produces coefficients which are integers or simple
fractions. 

At present it is not clear if the corrections that one finds by applying USEP to a field  automatically apply to, or are inherited by, other types of fields.
The suggested luminosity distance may eventually turn out to be
produced by considering other fields\footnote{It is interesting, for instance, that a
different mechanism, based on a Casimir-like vacuum
energy~\cite{urban}, needs a Veneziano ghost in order reproduce a
density of  the right order of magnitude.} than the scalar field considered in~\cite{fedo}  or by means of other theoretical insights. 
We also considered the full three parameters model (\ref{ODE})-(\ref{ODE2}), whose best fit considerably improves the $\chi^2$ and   is found to be favored over $\Lambda$CDM by the AIC but not by the BIC criterion.

It should be noted that the parameters that we are fitting are not coupling constants, and do not appear in a Lagrangian. Rather, they are intended as the terms of a series expansion that approximate the new ``IR-completed" theory starting from GR. As mentioned in the introduction, the proposed deformation is present at any time during the cosmological evolution; it 
affects any cosmological observation at high redshift, regardless of when such
an observation takes place, and therefore addresses the ``coincidence problem" in the most direct way.

We note that our model requires a rather low value of the
Hubble constant, $H_0 \sim 50$\,km/s/Mpc, compared to the
constraint from the Hubble Key Project with the determination of
$H_0=72 \pm 8$\,km/s/Mpc \cite{Freedman}. This comes from the fact
that the Hubble parameter evolves as $H(z) \simeq H_0 (1+z)^{3/2}$
even for $z \lesssim {\cal O}(1)$ due to the absence of a dark
energy component. However, methods of the determination of $H_0$
that are largely independent of distance scales of the Large
Magellanic Cloud Cepheid typically give low values of $H_0$
\cite{Sarkar}. For example, Reese {\it et al.} \cite{Reese} showed
that Sunyaev-Zel'dovich distances to 41 clusters provide the
constraint $H_0=54^{+4}_{-3}$\,km/s/Mpc in the Einstein de Sitter
universe. Thus the information of $H_0$ alone is not yet
sufficient to rule out our model.

It would be interesting to see the effect of the proposed IR correction on the
angular diameter distance to the last scattering surface and estimate the modification to the temperature anisotropies in
Cosmic Microwave Background (CMB). CMB data will be
certainly useful to place further constraints on our model and to complete the picture at higher redshift.

\section*{ACKNOWLEDGEMENTS}
F.\,P. is indebted to Bruce A.~Bassett for many valuable
conversations during his visit at Perimeter Institute. We
thank Justin Khoury for his suggestions on the manuscript. S.\,N.
acknowledges support from the Niels Bohr International Academy,
the EU FP6 Marie Curie Research $\&$ Training Network
``UniverseNet" under Contract No. MRTN-CT-2006-035863 and the
Danish Research Council under FNU Grant No. 272-08-0285. The research
of F.\,P. at Perimeter Institute is supported in part by the Government
of Canada through NSERC and by the Province of Ontario through the
Ministry of Research \& Innovation. S.\,T. thanks financial
support for JSPS (Grant No.~30318802).

\appendix
\section{USEP and the derivation of Eq.~(\ref{dotlamre})}
\label{sketch}

While referring to~\cite{fedo} for details and motivations, it is
worth summarizing here the basic steps that lead  to
Eq.~(\ref{dotlamre}), based on the ``ultra strong equivalence
principle" (USEP). The USEP is a statement about the bare energy
momentum tensor of the quantized fields on a general background,
namely:
\begin{quote}
USEP: \emph{For each matter field or sector sufficiently decoupled
from all other matter fields, there exists a state, the ``vacuum",
for which the expectation value of the (bare) energy momentum
tensor reads the same as in flat space, regardless of the
configuration of the gravitational field.}
\end{quote}

Our starting point is a free scalar field with the action
\begin{equation} \label{action}
S = \frac{1}{2}\int \rd^4 x \sqrt{-g}
\left(\partial \phi^2 - m^2 \phi^2\right)\,,
\end{equation}
in a spatially flat FLRW metric:
\begin{equation}
\rd s^2 = \rd t^2 - a^2(t)\rd \vec{x}^2\,.
\end{equation}
The equations of motions of the field and the $00$ component
of its energy momentum tensor read, respectively,
\begin{eqnarray}
\hspace{-2.5em}& &\ddot \phi(t, \vec{x}) + 3 H \dot \phi (t, \vec{x}) - \frac{\partial_i^2}{a(t)^2}  \phi (t, \vec{x}) + m^2 \phi (t, \vec{x}) = 0,
\label{equation}\\
\hspace{-2.5em}& &T^0_0(t, \vec{x})  = \frac{1}{2}\left[{\dot \phi}^2 + \frac{1}{a^2} (\partial_j \phi)^2 + m^2 \phi^2\right].
\label{calH}
\end{eqnarray}
In order to apply USEP we now review the standard calculation of
the energy momentum VEV in such a spacetime and compare it to the
flat space result.

Upon standard quantization in the Heisenberg picture the field is
expanded in creators and annihilators:
\begin{equation} \label{field}
\phi(t, \vec{x}) = \int \rd n^3\left[\psi_n(t) e^{i \vec{n} \cdot \vec{x}} A_{\vec{n}} + \psi^*_n(t) e^{-i \vec{n} \cdot \vec{x}} A^{ \dagger}_{\vec{n}}\right]\,,
\end{equation}
where $\vec{n}$ is the comoving momentum label in the FLRW space;
$\vec{n}$ is a conserved quantity, related to the proper physical
momentum $\vec{p}$ by $\vec{p} = \vec{n}/a(t)$. {}From the
canonical commutation relations $[\phi (\vec{x}), \pi(\vec{x}\,
')] = i(2\pi)^3 \delta^3(\vec{x}-\vec{x}\, ')$, the commutation
relations among the global Fourier operators are easily derivable,
\begin{equation} \label{normalcom}
[A_{\vec{n}}, A^{ \dagger}_{\vec{n} '}] = \delta^3(\vec{n} - \vec{n}\, ')\, .
\end{equation}
It is customary to choose $A_{\vec{n}}$ as the operator that
always annihilates the vacuum. The chosen vacuum state is
therefore implicitly characterized by the choice of the mode
functions $\psi_n(t)$, that, by (\ref{field}) and
(\ref{equation}), satisfy
\begin{equation} \label{modes0}
\ddot \psi_n + 3 H \dot \psi_n +  \omega_n^2 \psi_n = 0 ,
\end{equation}
where $\omega_n = \sqrt{n^2/a^2 + m^2}$.

The solutions of (\ref{modes0}) corresponding to the adiabatic
vacuum~\cite{birrell,fullings} can be found by a formal WKB
expansion. After quantization, the energy momentum tensor of the
field (\ref{calH}) becomes an operator whose expectation value on
the adiabatic vacuum reads~\cite{fullings,fedo}
\begin{equation} \label{square}
\langle T_0^0 (t, \vec{x})  \rangle = \frac{1}{4 \pi^2 a^3}
\int_0^\infty n^2 \left[\omega_n + \frac{H^2 a }{2 n} + {\cal O}(n^{-3})\right]
\,\rd n\,.
\end{equation}
The above should be compared to the flat space result
\begin{equation} \label{square2}
\langle T_0^0 (t, \vec{x}) \rangle_{\rm flat} = \frac{1}{4 \pi^2 a^3}
\int_0^\infty n^2\omega_n  \,\rd n \,.
\end{equation}

It is known that there is a strict connection between the
geometric properties of a manifold and the spectrum of the
differential operators \cite{heat} or the algebra of functions
\cite{connes} therein defined; such abstract characterizations
have occasionally been used for generalizing common geometrical
concepts and the description of spacetime itself~\cite{connes,sw}.
However, so far, attempts in this direction have always been
applied to the UV and intended to modify spacetime at the smallest
scales. Here we would like to modify the IR-spectral properties of
the FLRW metric (and therefore its geometry) in order to enforce
USEP. The proposed deformation is argued to correspond to a
breakdown of the metric Riemannian structure at distances
comparable to $H^{-1}$.

So the idea is to modify the physics in the infra-red but strictly
maintain the equations and the relations valid locally such as the
field equations (\ref{equation}) and the form of the energy
momentum tensor (\ref{calH}). We choose a point in FLRW ($\vec{x}
= 0$) and make a formal Taylor expansion of which GR is the zeroth
order. In the spirit of a general spectral deformation, we
conjecture a mismatch between the ``metric-manifold" Fourier
labels $\vec{n}$ and the physical momenta $\vec{k}$ that locally
define the infinitesimal translations and the derivatives of the
local fields. In other words, we now write the field in $\vec{x}
\simeq 0$ as
\begin{equation} \label{localfield}
\phi(t, \vec{x}\approx 0) = \int \rd n^3\left[\psi_k(t) e^{i \vec{k} \cdot \vec{x}} A_{\vec{n}}
+ \psi^*_k(t) e^{-i \vec{k} \cdot \vec{x}} A^\dagger_{\vec{n}}\right]\,,
\end{equation}
where
\begin{equation} \label{kk}
\vec{k} = \vec{n} \left(1 - \frac{H^2 a^2}{2 n^2} + {\rm higher\ order} \right)\,.
\end{equation}
Note that, when derivatives of the field are taken in $\vec{x}=0$,
a factor of $k$, instead of $n$, drops. The form of above
relation, which is one of the main results of~\cite{fedo}, is
dictated by the request that the quadratically divergent time
dependent piece of (\ref{square}) disappears. In other words, that
is the first order correction in order to impose USEP upon this
particular GR solution. The corrected  mode equation is in fact
obtained by substituting (\ref{localfield}) into (\ref{equation}),
which is assumed to be strictly valid because it applies locally.
To the modified mode equation, $\ddot{\psi}_k(t)  + 3 H
\dot{\psi}_k(t) +  \omega^2_k \psi_k(t) = 0$, the same WKB
expansion can be applied  and the quadratic divergence in
(\ref{square}) is reabsorbed just by re-expressing $\langle 0 |\,
T_0^0  | 0 \rangle$ in terms of the appropriate ``flat measure"
time-independent Fourier coordinates $n$:
\begin{eqnarray} \nonumber
\langle T_0^0 (t, \vec{x}\approx 0) \rangle \!\!&=& \!\! \frac{1}{4 \pi^2 a^3}
\int_0^\infty n^2 \left[\omega_k + \frac{H^2 a }{2 k} + {\cal O}(n^{-3})\right]\,\rd n \\
& = & \frac{1}{4 \pi^2 a^3} \nonumber
\int_0^\infty n^2 \left[\omega_n  + {\cal O}(n^{-3})\right]\, \rd n\,.
\end{eqnarray}

In order to define local quantities away from the origin we
exploit the assumption of spatial homogeneity and use the
translation operator
\begin{equation}
T(\vec{\lambda}) = \exp \left({-i \vec{\lambda} \cdot \int d^3n\, \vec{k}(\vec{n})\,
A^\dagger_{\vec{n}} A_{\vec{n}}}\right)\,,
\label{translation}
\end{equation}
that we obtain by simple exponentiation of the (modified!)
momentum  operator. The parameter $\lambda$ is the comoving
distance. The field away from $\vec{x}=0$ is thus defined as
$\phi(t, \vec{\lambda}) \equiv T_i(\lambda)\, \phi(t, 0) \,
T_i^{-1}(\lambda)$ and reads
\begin{equation} \label{localfar}
\phi(t, \vec{\lambda}) =
 \int \rd^3n\,  \left[\psi_k(t) e^{i \vec{k} \cdot \vec{\lambda}} A_{\vec{n}} + \psi^*_k(t) e^{-i \vec{k} \cdot \vec{\lambda}} A^\dagger_{\vec{n}}\right] .
\end{equation}
{}From the above expression it is straightforward to calculate the
modified  commutator between the canonical momentum $\pi(0) = a^3
{\dot \phi}(0)$ and the field at comoving distance
$\vec{\lambda}$,
\begin{equation} \label{comm}
[\pi(0), \phi( \vec{\lambda})] = -i\left(\delta^3(\vec{\lambda}) + \frac{1}{8 \pi} \frac{H^2 a^2}{\lambda} \right) .
\end{equation}
Note that there is a potential ambiguity in defining the time
derivative of a displaced operator. By deriving (\ref{localfar})
we get
\begin{eqnarray}
 \pi(\vec{\lambda}) \!\!&=& \!\! a^3 \nonumber
 \int \rd^3n\,  \left[\dot{\psi}_k(t) e^{i \vec{k} \cdot \vec{\lambda}} A_{\vec{n}} + \dot{\psi}^*_k(t) e^{-i \vec{k} \cdot \vec{\lambda}} A^\dagger_{\vec{n}}\right] \\
 &+& i a^3  \int \rd^3n\,  (\vec{k} \cdot \vec{\lambda})\dot{}\ \left[\psi_k  \ e^{i \vec{k} \cdot \vec{\lambda}} A_{\vec{n}}  - \psi^*_k  \ e^{- i \vec{k} \cdot \vec{\lambda}} A^\dagger_{\vec{n}} \right]. \nonumber
\end{eqnarray}
The second line in the above equation is there because $k$ is time
dependent. However, if we just apply the translation to $\pi(0)$,
instead of deriving $\phi(\vec{\lambda})$, those terms would not
be there. Therefore, for consistency, we need to make them
ineffective at the required order of approximation. This can be
done by imposing that  $[\phi(0), \pi(\vec{\lambda})] = -
[\pi(0),\phi(\vec{\lambda})]$. Because of the second line the last
equation, the commutator between $\phi(0)$ and $\pi(\lambda)$
gives
\begin{eqnarray} \label{nonsym}
[\phi(0), \pi(\vec{\lambda})] &=& - [\pi(0),\phi(\vec{\lambda})] \nonumber \\
&&- 2i \frac{a^3}{(2\pi)^3}
\int \rd^3 n \, e^{- i \vec{k} \cdot \vec{\lambda}}
\left| \psi_n \right|^2 (\vec{k} \cdot \vec{\lambda})\dot{}\,,
\nonumber \\
\end{eqnarray}
the last term being the spurious contribution. In order to get rid
of it, we have to make the comoving physical distance
$\vec{\lambda}$ also time dependent. This effectively means that,
after an infinitesimal time step $\rd t$, we have to reconsider
the field translated, from  $\vec{x} \approx 0$, not by the same
comoving distance $\lambda$, but by a slightly different amount.

At high momenta/small distances, since $|\psi_k|^2 \sim 1/n$, the
integral in the second term of (\ref{nonsym}) reads
$$
\int \rd^3 n \, e^{- i \vec{n}  \cdot \vec{\lambda}} \, \frac{1}{n} \left[\dot{\vec{\lambda}} \cdot \vec{n}\left(1 - \frac{H^2 a^2}{2 n^2}\right) - \vec{\lambda}\cdot \vec{n} \frac{(H^2 a^2)\dot{}}{2 n^2}\right]\, .\nonumber
$$
We make the ansatz $\dot{\vec{\lambda}} = b \lambda^2 (H^2 a^2)\dot{}\, \vec{\lambda}$,
where $b$ is a number to be determined. We get
\begin{eqnarray} & & \nonumber
\int \rd^3 n \, e^{- i \vec{k} \cdot \vec{\lambda}} \left| \psi_n \right|^2 (\vec{k} \cdot \vec{\lambda})\dot{}\ \\ & \simeq & \ i  (H^2 a^2)\dot{} \ \frac{ d}{ d \alpha} \int d^3 n\, \left. \left(\frac{b \lambda^2}{n} - \frac{1}{2 n^3}\right) e^{- i \alpha\, \vec{n} \cdot \vec{\lambda}}\right|_{\alpha =1}. \nonumber
\end{eqnarray}
The last integral can be regularized by setting $n^{-3}
\rightarrow n^{-3 + \epsilon}$ and taking the $\epsilon
\rightarrow 0$ limit only after deriving with respect to $\alpha$.
The result is null for $b = 1/4$, which fixes the time dependence
of $\lambda$:
\begin{equation} \label{expansion}
\dot{\lambda} = \lambda^3 \frac{(H^2 a^2)\dot{}}{4} .
\end{equation}

\section{Exact solutions}
\label{appendixsec}

Here we present some analytical solutions for the differential
equation (\ref{ODE}). If $\gamma=0$ we obtain the following
analytic solution:
\begin{equation}
H_0 r(z)=\frac{4/\sqrt{1+z}}{-1-2 \alpha +\sqrt{\delta}+
2\sqrt{\delta }/[(1+z)^{\sqrt{\delta}/2}-1]}\,,
\end{equation}
where $\delta=(1+2\alpha)^2-16\beta$.

When $\gamma \neq 0$, finding the solution is much more difficult
and can only be given in an implicit form. For example, let us
consider the correction in Eq.~(\ref{redshift}), i.e.
$\alpha=\beta=0$ and $\gamma=1/4$. Setting $R(z)\equiv r(z)
\sqrt{1+z}$, we get a differential equation for the function
$R(z)$:
\begin{equation}
\frac{\rd R(z)}{\rd z}=\frac{1}{1+z}+
\frac{R(z)}{2(1+z)}+\frac{R(z)^3}{4 (1+z)}\,,
\label{Rzeq}
\end{equation}
with initial conditions $R(0)=0$ and $(\rd R/\rd z)(0)=1$.
By direct differentiation it can be shown that the solution
to Eq.~(\ref{Rzeq}) is given in an implicit form by:
\begin{eqnarray}
z &=& -1+\left(\frac{R(z)}{x_1}-1\right)^{\frac{4}{2+3 x_1^2}}
\left(\frac{R(z)}{x_2}-1\right)^{\frac{4}{2+3 x_2^2}}
\nonumber \\
& &\,\,\,\,\,\,\,\,\,
\times \left(\frac{R(z)}{x_3}-1\right)^{\frac{4}{2+3 x_3^2}}\,,
\end{eqnarray}
where the parameters $x_1,x_2,x_3$ are
the roots of the polynomial equation:
\begin{equation}
4+2 x_i+x_i^3=0\,.
\end{equation}

When all three parameters $\alpha$, $\beta$, and $\gamma$
are not zero, then we can still find an implicit solution for
$r(z)$ but in this case it is very complicated, so we will not
reproduce it here.

\end{document}